\documentclass[prd,showpacs,preprintnumbers,amsmath,amssymb]{revtex4}
\usepackage{graphicx}% Include figure files
\usepackage{dcolumn}% Align table columns on decimal point
\usepackage{bm}% bold math

\begin{document}

\title{Construction of Exact Ermakov-Pinney Solutions and Time-Dependent Quantum Oscillators}

%\author{Eun-Joo Kang}
\author{Sang Pyo Kim}
\email{sangkim@kunsan.ac.kr}
\author{Won Kim}
\affiliation{Department of Physics, Kunsan National University, Kunsan 54150,
 Korea}

\begin{abstract}
The harmonic oscillator with a time-dependent frequency has a family of linear quantum invariants for the time-dependent Schr\"{o}dinger equation, which are determined by any two independent solutions to the classical equation of motion. Ermakov and Pinney have shown that a general solution to the time-dependent oscillator with an inverse cubic term can be expressed in terms of two independent solutions to the time-dependent oscillator. We explore the connection between linear quantum invariants and the Ermakov-Pinney solution for the time-dependent harmonic oscillator. We advance a novel method to construct Ermakov-Pinney solutions to a class of time-dependent oscillators and the wave functions for the time-dependent Schr\"{o}dinger equation. We further show that the first and the second P\"{o}schl-Teller potentials belong to a special class of exact time-dependent oscillators. A perturbation method is proposed for any slowly-varying time-dependent frequency.
\end{abstract}

\date{\today}
\pacs{03.65.Ge, 03.65.-w, 45.20.-d, 05.45.-a\\Keywords: Time-dependent oscillator, Ermakov-Pinney solution, Quantum invariant, P\"{o}schl-Teller potential}

\maketitle

\section{Introduction}\label{sec1}

The harmonic oscillator with a time-dependent frequency has been studied as a classical or quantum model for a dissipative or nonequilibrium system. The Lorentz's pendulum with a varying length is a simple time-dependent oscillator, whose classical adiabatic invariants were investigated by Chandrasekhar \cite{chandrasekhar} and Littlewood \cite{littlewood}. Each Fourier mode of a massive linear field in a homogeneous, time-dependent spacetime \cite{guth-pi85} or a charged scalar field in a homogeneous, time-dependent electric field is characterized by a harmonic oscillator with a time-dependent frequency \cite{kim14b}. The charged scalar field in a time-dependent, homogeneous magnetic field is equivalent to that of an infinite system of coupled oscillators for Landau levels \cite{kim14a}. The quantum theory of time-dependent oscillators provides shortcuts to adiabaticity with an important, analytical model \cite{crscgm,deffner-jarzynski-campo}.

Long before the advent of quantum theory, Ermakov studied the integrability of a nonlinear system with a time-dependent harmonic force with an inverse cubic term, which is now known as the Ermakov invariant \cite{ermakov80} (for a review, see Ref. \citenum{leach-andriopoulos}). It was Pinney who found a general solution to the time-dependent oscillator with an inverse cubic term  in terms of two independent solutions for the time-dependent harmonic oscillator \cite{pinney50}. Lewis and Riesenfeld introduced a quadratic invariant for a time-dependent oscillator, whose invariant satisfied the quantum Liouville equation and whose eigenstate provided an exact solution to the time-dependent Schr\"{o}dinger equation \cite{lewis-riesenfeld69}. A pair of quantum invariants linear in the momentum and the position operators was introduced by Malkin, Man'ko and Trifonov \cite{malkin-manko-trifonov70,malkin-manko-trifonov73}, and independently by one (SPK) of the authors of this paper \cite{kim95,kim-kim99} and also in Ref. \citenum{ffvv98}. In fact, the quantum invariant pair play roles as the time-dependent annihilation and creation operators for the time-dependent oscillator exactly in the same way as for a static oscillator \cite{kim-lee00,kim-page01}.

In this paper, we explore the connection between the linear quantum invariants and the Ermakov-Pinney solution for a time-dependent harmonic oscillator. The general Ermakov-Pinney solution to the time-dependent oscillator, whose invariant is given by two independent solutions, can be expressed in terms of a complex solution, which leads to a pair of quantum invariant operators playing the roles of time-dependent annihilation and creation operators. The quantization condition determines the phase of the complex solution as a function of the amplitude. This implies that exact time-dependent oscillators can be constructed from the form of the amplitude. We then advance a novel method to find the solutions to the Ermakov-Pinney solutions and the exact time-dependent oscillators. We investigate time-dependent oscillators whose amplitudes for complex solutions are exponentially varying, power-law varying, and oscillating. The hyperbolic and the sinusoidal amplitudes recover the first and the second P\"{o}schl-Teller potentials. Finally, we propose a perturbation method to systematically find solutions for any slowly-varying time-dependent frequency.

The organization of this paper is as follows. In Section \ref{sec2}, we revisit the Ermakov-Pinney solution to a time-dependent oscillator with an inverse cubic term and then show the connection between the Ermakov-Pinney solution and the complex solution for quantization of the time-dependent oscillator. In Section \ref{sec3}, we use the complex solution to find a pair of linear quantum invariant operators for a time-dependent oscillator, these invariant operators playing the roles of time-dependent annihilation and creation operators. In Section \ref{sec4}, we advance a method to construct the exact Ermakov-Pinney solutions and the quantum invariant operators for time-dependent oscillators and classify the time-dependent frequencies, which exhibit exponential, power-law, oscillatory behaviors and include the first and the second P\"{o}schl-Teller potential. In Section \ref{sec5}, we propose a perturbation method and compare its results with the higher-order WKB (Wentzel-Kramers-Brillouin) results.

\section{Ermakov-Pinney Solution}\label{sec2}

More than a century ago Ermakov studied the integrability of the following nonlinear differential equation \cite{ermakov80}:
\begin{eqnarray} \label{ermakov}
\ddot{x} (t) +\omega^{2} (t) {x} (t) = \frac{L^2}{x^3 (t)},
\end{eqnarray}
where the last term may come from the angular momentum when $x$ denotes the radial coordinate in two or higher dimensions under a time-dependent central force $- \omega^2 (t) x$.
A Hamiltonian leading to the Ermakov equation may be given by using a time-dependent oscillator with an inverse square potential:
\begin{eqnarray} \label{time ham}
H_L (t) = \frac{p^2}{2} + \frac{\omega^2 (t)}{2} q^2 + \frac{L^2}{2 q^2}.
\end{eqnarray}
Pinney found a general solution to Eq. (\ref{ermakov}), which is given by \cite{pinney50}
\begin{eqnarray} \label{pinney}
x(t)=(Au^{2} (t) +2B u (t) v (t) +Cv^{2} (t) )^{1/2},
\end{eqnarray}
where $AC \geq B^2$ and $u$ and $v$ are two independent solutions to the equation for a linear oscillator
\begin{eqnarray} \label{osc eq}
\frac{d^2}{dt^2}\binom{u (t) }{v (t)}+\omega^{2} (t) \binom{u (t)}{v(t)}=0. \label{lin eq}
\end{eqnarray}
Noting that
\begin{eqnarray}
x\dot{x} = Au\dot{u}+B(v\dot{u}+u\dot{v})+C v\dot{v}
\end{eqnarray}
and using Eq. (\ref{osc eq}), we can show that
\begin{eqnarray}
x(\ddot{x}+\omega^{2}x ) =\frac{({\rm Wr}[u,v])^{2}(AC-B^{2})}{x^{2}}.
\end{eqnarray}
Finally, by choosing the constants such that
\begin{eqnarray} \label{er con}
AC-B^{2}=\Bigl( \frac{L}{{\rm Wr} [u,v]} \Bigr)^2,
\end{eqnarray}
we show that Eq. (\ref{pinney}), indeed, satisfies the so-called Pinney equation, Eq. (\ref{ermakov}).

On the other hand, the quantum theory for a linear oscillator $H_0 (t)$ may be considered in connection to the nonlinear oscillator $H_L (t)$.
For that purpose, we may introduce a pair of complex solutions $w$ and $w^*$ for $H_0 (t)$ as
\begin{eqnarray}
Au^{2}+2Buv+Cv^{2}= w w^*,
\end{eqnarray}
where
\begin{eqnarray}
w = \sqrt{A}u+\frac{B-i\sqrt{AC-B^{2}}}{\sqrt{A}}v, \quad w^* = {\rm c. c.};
\end{eqnarray}
then, we may impose another Wronskian condition (in unit of $\hbar =1$)
\begin{eqnarray} \label{wr con}
{\rm Wr}[w, w^*] = 2i\sqrt{AC-B^{2}} {\rm Wr} [u, v] = i.
\end{eqnarray}
The Wronskian  condition, Eq. (\ref{wr con}), is equal to $L =1/2$ and is necessary for the quantization rule, which will be used to introduce linear quantum invariants in the next section. Setting
\begin{eqnarray}
w (t) = \xi (t) e^{-i\theta (t)}, \quad \xi = \sqrt{w^* w},
\end{eqnarray}
where $w$ is a complex solution to Eq. (\ref{osc eq}), and using the Wronskian condition in Eq. (\ref{wr con})
\begin{eqnarray} \label{phase}
2 \xi^{2} \dot{\theta}=1,
\end{eqnarray}
we find the equation for $\xi$
\begin{eqnarray} \label{lr eq}
\ddot{\xi}+\omega^{2}\xi= \frac{1}{4\xi^{3}}.
\end{eqnarray}
Then, the complex solution $\xi$ is given by
\begin{eqnarray}
w (t) = \xi (t) e^{- i \int^{t} \frac{ dt'}{2 \xi^2 (t')}}. \label{pol sol}
\end{eqnarray}
Note that $x = \sqrt{2} \xi$ and that $G = x^2$ satisfies the linear equation discovered by Gel'fand and Dikii \cite{gelfand-dikii75}.

\section{Connection to Quantum Invariants} \label{sec3}

For the time-dependent Schr\"{o}dinger equation for the Hamiltonian $H_0 (t)$ with $L = 0$, Lewis and Riesenfeld found a quadratic invariant operator \cite{lewis-riesenfeld69}
\begin{eqnarray} \label{lr in}
\hat{I}_0 (t) =\frac{1}{2}\Bigl( (\xi \hat{p} - \dot{\xi} \hat{q} )^{2}+\bigl(\frac{\hat{q}}{\xi} \bigr)^{2} \Bigr),
\end{eqnarray}
where $\xi$ satisfied Eq. (\ref{lr eq}). In fact, $\hat{I}_0$ is a solution to the quantum Liouville equation
\begin{eqnarray}
i \frac{\partial \hat{I}_0 (t)}{\partial t} + [\hat{I}_0 (t), \hat{H}_0 (t) ] = 0.
\end{eqnarray}
The quadratic invariant for the Hamiltonian $\hat{H}_L (t)$ for $L \neq 0$ was used to construct the coherent and the squeezed states in Ref. \citenum{kim-kim95}.

We now find the connection between the Ermakov-Pinney solutions and the quantum states constructed by using a pair of linear invariants
\begin{eqnarray}
\hat{a} (t) = i(w^* \hat{p} -\dot{w}^* \hat{q} ), \quad \hat{a}^{\dagger} (t) = - i(w \hat{p}-\dot{w} \hat{q}),
\end{eqnarray}
where $\hat{p}$ and $\hat{q}$ are the operators in the Schr\"{o}dinger picture. The operators $\hat{a} (t)$ and $\hat{a}^{\dagger}(t)$ act as the time-dependent annihilation and creation operators in the interaction picture.
We can show a new quadratic invariant of the number operator:
\begin{eqnarray} \label{ham in}
\hat{I}_0 &=& \omega_0 \Bigl(\hat{a}^{\dagger} \hat{a} +\frac{1}{2} \Bigr) \nonumber\\
&=& \frac{\omega_0}{2} \Bigl( \bigl( \xi \hat{p}-\dot{\xi} \hat{q} \bigr)^{2}+\bigl(\frac{\hat{q}}{\xi} \bigr)^{2} \Bigr).
\end{eqnarray}
Note that the invariant becomes the Lewis-Riesenfeld invariant when $\omega_0 =1$. The ground state and the excited number states are given by
\begin{eqnarray} \label{num st}
\hat{a} (t) \vert 0, t \rangle = 0, \quad \vert n, t \rangle = \frac{\bigl(\hat{a}^{\dagger} (t) \bigr)^n}{\sqrt{n!}} \vert 0, t \rangle.
\end{eqnarray}
The wave function of the number state in Eq. (\ref{num st}) is explicitly given in Eq. (52) of Ref.~\citenum{kim-page01}, which in terms of $\xi$ reads
\begin{eqnarray} \label{wav fn}
\Psi_n (q, \xi (t)) = \frac{1}{\sqrt{(2 \pi)^{1/2} 2^n n! \xi (t) }} e^{- \frac{i}{2} \bigl(n+ \frac{1}{2} \bigr) \int^t \frac{dt'}{\xi^2 (t')} } H_n \Bigl( \frac{q}{\sqrt{2} \xi (t)} \Bigr) e^{- \bigl( \frac{1}{4 \xi^2 (t)} - i \frac{\dot{\xi} (t)}{2 \xi (t)} \bigr) q^2 },
\end{eqnarray}
where $H_n$ is the Hermite polynomial. The wave function in Eq. (\ref{wav fn}) is the same as Eq.~(58) of Ref.~\citenum{dabrowski-dunne}.

We may introduce a new parametrization of constants as
\begin{eqnarray}
A = D \cosh re^{-i\varphi}, \quad B=D\sinh r, \quad C = D \cosh re^{i\varphi},
\end{eqnarray}
and express the solution as
\begin{eqnarray}
w =\frac{\sqrt{D}}{\sqrt{\cosh r}} \bigl( \cosh re^{-i\varphi/2}u +(\sinh r-i)v \bigr).
 \end{eqnarray}
The dispersion relations are
\begin{eqnarray} \label{un rel}
\Delta p^{2}=\dot{w}^*\dot{w}, \quad \Delta q^{2} = w^* w, \quad \bigl( \Delta q \Delta p \bigr)^{2} = (\xi \dot{\xi})^{2}+\frac{1}{4}.
\end{eqnarray}
The minimal uncertainty is given by the condition
\begin{eqnarray}
\dot{\xi}=0,
\end{eqnarray}
i.e., when $\xi = {\rm constant}$. Owing to Eq. (\ref{phase}), the solution $w_0 = \xi_0 e^{- i t/(2 \xi_0^2)}$ for a constant $\xi_0$ provides a minimal uncertainty state, and the coherent states constructed from $w_0$ have the same minimal uncertainty.

\section{Construction of Ermakov-Pinney Solutions} \label{sec4}

We note that $\xi \dot{\xi}$ measures not only the uncertainty relation in Eq. (\ref{un rel}) but also the expectation value
\begin{eqnarray}
\langle 0, t \vert \hat{H}_0 \vert 0, t \rangle = \frac{1}{2} \dot{\xi}^2 + \frac{\omega^2(t)}{2} \xi^2 + \frac{1}{8 \xi^2}.
\end{eqnarray}
In order to use the Ermakov-Pinney solution in finding the exact the time-dependent oscillators, we introduce a measure for the variation of amplitude:
\begin{eqnarray} \label{am meas}
\frac{d \xi^{2}(t)}{dt} = \epsilon(t).
\end{eqnarray}
Then, using Eqs.~(\ref{lr eq}) and (\ref{am meas}), we may determine the frequency:
\begin{eqnarray} \label{freq}
\omega^{2} (t) = \frac{1+ \epsilon^{2} - 2\dot{\epsilon} \xi^{2}}{4\xi^{4}}. \label{con fr}
\end{eqnarray}
The characteristic behavior of $\xi$, and thereby $\epsilon$, determines the frequency in Eq. (\ref{freq}) and the Ermakov-Pinney solution in Eq. (\ref{ermakov}). Below we classify some exact models not only for the Ermakov-Pinney solution but also the corresponding quantum oscillator.

\subsection{Exponential Behavior}
As the first model, we consider an exponentially-varying amplitude $\xi^2 = \epsilon_0 e^{\lambda t}/\lambda $ with $\epsilon=\epsilon_0 e^{\lambda t}$ for $\lambda \neq 0$ and $\epsilon > 0$, which leads to the frequency
\begin{eqnarray}\label{fr-exp}
\omega^{2} (t) = \Bigl(\frac{\lambda}{2\epsilon_0} \Bigr)^{2} e^{2 \lambda t} - \frac{\lambda^{2}}{4}.
\end{eqnarray}
For an exponentially-growing (decreasing) amplitude in late (early) time, the frequency crosses the zero at $t_0 = - \ln(\epsilon_0)/\lambda$ and the characteristic behavior changes from an unstable (stable) motion to a stable (unstable) one due to the change in sign.

The second exponential behavior is provided by $\xi^2 = \epsilon_0 \tau \cosh(t/\tau) $ and $\epsilon=\epsilon_0 \sinh (t/\tau)$ for $\tau \neq 0$ and $\epsilon > 0$. Then, the corresponding frequency is
\begin{eqnarray}\label{fr-cosh}
\omega^{2} (t) = \frac{1- \epsilon_0^2}{4(\epsilon_0 \tau)^2} \frac{1}{\cosh^2 (t/\tau)} - \frac{1}{4 \tau^2}.
\end{eqnarray}
Another behavior is given by $\xi^2 = \epsilon_0 \tau \sinh(t/\tau) $ and $\epsilon=\epsilon_0 \cosh (t/\tau)$ with the frequency
\begin{eqnarray}\label{fr-sinh}
\omega^{2} (t) = \frac{1 + \epsilon_0^2}{4(\epsilon_0 \tau)^2} \frac{1}{\sinh^2 (t/\tau)} - \frac{1}{4 \tau^2}.
\end{eqnarray}
In quantum mechanics, the frequency in Eq. (\ref{fr-cosh}) or Eq. (\ref{fr-sinh}) corresponds to the second P\"{o}schl-Teller potential with a specific energy \cite{poschl-teller}. The algebraic structure of the second P\"{o}schl-Teller potential was studied in Ref.~\citenum{barut-inomata-wilson87}.

\subsection{Power-Law Behavior}
The power-law behavior is given by $\xi^2 = \tau \epsilon_0/(n+1) (t/\tau)^{n+1}$ and $\epsilon (t) =\epsilon_0 (t/\tau)^{n}$ for $n \neq -1$, which leads to the frequency
\begin{eqnarray}
\omega^{2} (t) = \frac{(n+1)^2 + \epsilon_0^2 (1-n) (n+1)^2 (\frac{t}{\tau})^{2n}}{4 (\epsilon_0 \tau)^{2} (\frac{t}{\tau})^{2n+2}}.
\end{eqnarray}
Note the cross-over behavior for $n > 1$ from a stable motion to an unstable one. No cross-over behavior is seen for $n \leq 1$. The positive frequency solution to Eq.~(\ref{osc eq}) is given by the Hankel function
\begin{eqnarray}
w (t) = \frac{\sqrt{\pi t}}{2} H_{\frac{n^2 + n - 1}{4n}} \Bigl( \frac{n+1}{2 \epsilon_0 \tau n} t \Bigr).
\end{eqnarray}
In the case of $n = 0$, the frequency takes the form
\begin{eqnarray}
\omega^{2} (t) =\frac{1 + \epsilon_0^{2}}{4(\epsilon_0 t)^{2}}.
\end{eqnarray}
Then, the positive frequency solution is
\begin{eqnarray}
w = \sqrt{\frac{\epsilon_0 t}{\tau}}e^{-i (\tau/2\epsilon_0 ) \ln t},
\end{eqnarray}
which has the uncertainty relation
\begin{eqnarray}
(\Delta q \Delta p)^{2}=\frac{1}{4}+\frac{\epsilon_0^{2}}{4}.
\end{eqnarray}
In the special case of $n=-1$ for $\xi^2 = \epsilon_0 \tau \ln (t)$ and $\epsilon= \epsilon_0 (\tau/t)$, the frequency becomes
\begin{eqnarray}
\omega^{2} (t) = \frac{t^2 + (\epsilon_0 \tau)^2 + 2 (\epsilon_0 \tau)^2 \ln (t)}{4 (\epsilon_0 \tau t)^{2} (\ln(t))^2}.
\end{eqnarray}
A cross-over time for the behavior of oscillators from an unstable motion to a stable one is seen.

\subsection{Oscillatory Behavior}
The final model is provided by an oscillating amplitude
$\xi^2 = a + b \cos(2 \omega_0 t)$ and $\epsilon = -2b \omega_0 \sin (2 \omega_0 t)$. The corresponding frequency is
\begin{eqnarray} \label{fr-osc}
\omega^2 (t) = \frac{1 - 4(a^2- b^2) \omega_0^2}{4 \bigl(a + b \cos(2 \omega_0 t) \bigr)^2} +  \omega_0^2.
\end{eqnarray}
Another form $\xi^2 = a + b \sin(2 \omega_0 t)$ and $\epsilon = 2b \omega_0 \cos (2 \omega_0 t)$ gives the same form as in Eq. (\ref{fr-osc}) with $\cos (2 \omega_0 t)$ being replaced by $\sin (2 \omega_0 t)$.
Note that a linear superposition of $u = e^{- i \omega_0 t}/\sqrt{2 \omega_0}$ and $u^*$ gives such an oscillatory behavior.
The quantum-mechanical problem corresponding to the frequency in Eq. (\ref{fr-osc}) with $a = 0$ is the first P\"{o}schl-Teller potential \cite{poschl-teller}. The algebraic structure of the first P\"{o}schl-Teller potential was studied in Ref.~\citenum{quesne}.

\section{Perturbation Method for A General Time-dependent Frequency} \label{sec5}

A question may be raised whether the construction method in Section \ref{sec4} can be applied to a general time-dependent frequency $\omega(t)$. This question is equivalent to solving explicitly Eq. (\ref{lin eq}) for any $\omega (t)$, which goes beyond the theory of linear differential equations. We now propose a perturbation method and compare its results with the higher-order WKB results for a quantum-mechanical system under the mapping $t =x$.

Equation (\ref{con fr}) for constructing frequency may be written as
\begin{eqnarray}
\Bigl( \frac{1}{2 \xi^2} \Bigr)^2 = \omega^2 - f(\xi^2; \epsilon, \dot{\epsilon}), \label{mas eq}
\end{eqnarray}
where $\xi$ is characterized by the magnitude of $\omega$ and
\begin{eqnarray}
f(\xi^2; \epsilon, \dot{\epsilon}) = \frac{\epsilon^{2} - 2\dot{\epsilon} \xi^{2}}{4\xi^{4}}
\end{eqnarray}
is characterized by the time scale of the variation of $\omega$.
From the polar form in Eq. (\ref{pol sol}), we may identify $\dot{\theta} = 1/2 \xi^2$ of Eq. (\ref{mas eq}) with the WKB action $S(x)$ for the quantum-mechanical system. We propose the following iterative scheme:
\begin{eqnarray}
\Bigl( \frac{1}{2 \xi_{(n)}^2} \Bigr)^2 = \omega^2 - f(\xi_{(n-1)}^2; \epsilon_{(n-1)}, \dot{\epsilon}_{(n-1)}), \quad n = 0, 1, \cdots. \label{per eq}
\end{eqnarray}
Here, we understand that $\xi_{(-1)} = \epsilon_{(-1)} = \dot{\epsilon}_{(-1)} = 0$. The leading approximation
\begin{eqnarray}
\xi_{(0)} = \frac{1}{\sqrt{2 \omega}}
\end{eqnarray}
gives the WKB result. The next improved approximation
\begin{eqnarray}
\frac{1}{2 \xi_{(1)}^2} = \omega + \frac{3\dot{\omega}^2}{8 \omega^3} - \frac{\ddot{\omega}}{4 \omega^2}
\end{eqnarray}
is the sum of the zeroth- and the second-order WKB results \cite{bender-orszag}. The structure of the perturbation series and its comparison with that of the WKB method \cite{bender-orszag}, and the phase-integral method \cite{froman-froman} will be addressed in a future publication.

\section{Conclusion}\label{sec5}

We have studied the connection between the Ermakov-Pinney solution for a time-dependent oscillator with an inverse cubic term and the linear quantum invariants for the time-dependent oscillator. A general solution to the Pinney equation has been found in terms of two independent solutions to the time-dependent oscillator. We have shown that a complex solution from the two independent solutions leads to two linear invariant operators for the time-dependent Schr\"{o}dinger equation, these operators playing the roles of time-dependent annihilation and creation operators. Further, we have introduced a novel method to construct the Ermakov-Pinney solutions for exact time-dependent oscillators and investigated the time-dependent harmonic oscillator models whose amplitudes of the complex solutions exhibit an exponential or power-law or oscillating behavior. In particular, the hyperbolic and the sinusoidal amplitudes recover the first and the second P\"{o}schl-Teller potentials.

An open issue is to show whether or not all known models for time-dependent oscillators or the corresponding quantum problems in configuration space can be constructed through this method, and whether new models beyond those in the literature can be found. A comparison of the perturbation method of this paper with the WKB method and the phase-integral method, a quest for new physics, and an elaboration of the mathematical theory will be subjects to be addressed in future publications.

\acknowledgments
SPK would like to thank Christian Schubert for useful discussions and Adolfo del Campo for providing shortcuts to adiabaticity. The authors thank the anonymous referee for suggesting approximation methods.
This research was supported by the Basic Science Research Program through the National Research Foundation of Korea (NRF) funded by the Ministry of Education (NRF-2015R1D1A1A01060626).

\end{document}